\documentclass[superscriptaddress,amsfonts,amssymb,amsmath,twocolumn,floatfix]{revtex4}

\usepackage[USenglish]{babel}
\usepackage{epsfig}
\usepackage{graphics}
\usepackage{graphicx}
\usepackage{amsmath}
\usepackage{color}
\usepackage{comment}
\usepackage{slashed}
\usepackage{siunitx}
\usepackage{hyperref}

\usepackage{soul}
\usepackage[normalem]{ulem}
\usepackage{xcolor}
\usepackage[makeroom]{cancel}

\hypersetup{
    colorlinks=true,
    linkcolor=blue,
    citecolor=blue,
    filecolor=magenta,
    urlcolor=cyan,
    pdfpagemode=FullScreen,
    }

\date{\today}

\begin{document}

\title{Probing the Schr{\"o}dinger-Newton equation in a Stern-Gerlach-like experiment}

\author{Gabriel H. S. Aguiar}
\email{ghs.aguiar@unesp.br}
\affiliation{Instituto de Física Teórica, 
Universidade Estadual Paulista, Rua Dr. Bento Teobaldo Ferraz, 271, 01140-070, São Paulo, São Paulo, Brazil}

\author{George E. A. Matsas}
\email{george.matsas@unesp.br}
\affiliation{Instituto de Física Teórica, Universidade Estadual Paulista, Rua Dr. Bento Teobaldo Ferraz, 271, 01140-070, São Paulo, São Paulo, Brazil}

\pacs{}

\begin{abstract}
      Explaining the behavior of macroscopic objects from the point of view of the quantum paradigm has challenged the scientific community for a century today. A mechanism of gravitational self-interaction, governed by the so-called Schr{\"o}dinger-Newton equation, is among the proposals that aim to shed some light on it. Despite all efforts, this mechanism has been proven difficult to probe. Here, we consider a Stern-Gerlach-like experiment to try it out. The Schr{\"o}dinger-Newton equation can be analytically solved under certain proper conditions, and a change-of-phase effect induced by the gravitational self-interacting potential can be calculated.
\end{abstract}

\maketitle

\section{Introduction}

According to nonrelativistic quantum mechanics, a localized free particle spreads over time depending on its mass~$m$ and wave-packet width~$\sigma$. Quantum mechanics implies that the combination of~$m$ and~$\sigma$ for free macroscopic objects would lead to a fast-spreading of the wave packet~\cite{diosi84}, which has not been observed till now (possibly because of the difficulty of isolating them from their environment). In addition, we have never observed spatial quantum-mechanical superposition of macroscopic objects. The debate on whether the no observation of these effects is a momentary difficulty posed by technology or an intrinsic impediment raised by nature (stemming from the Planck scale) is ongoing.

Among those who believe that quantum mechanics must be amended to describe the classical world are Di{\'o}si and Penrose~\cite{diosi84, penrose96, penrose98}. The mechanism of gravitational self-interaction idealized by the first aims to provide some explanation for the localization of macroscopic objects. According to this proposal, a nonrelativistic quantum particle would be ruled by the Schr{\"o}dinger-Newton equation, which incorporates a gravitational self-interacting potential to the usual Schr{\"o}dinger equation. As originally discussed by Di{\'o}si, the gravitational self-interaction would prevent wave packets of sufficiently massive particles from spreading and staying in superposition (although he already raised issues with the Schr{\"o}dinger-Newton equation in the case of two-rotating solitons~\cite{diosi84}). Despite the Schr{\"o}dinger-Newton equation needing an additional mechanism to cope with the wave function {\em ``collapse''}~\cite{diosi84, diosi22}, it continues to be widely investigated, since it serves (at least) as a reference to quantitatively compare the predictions of usual quantum mechanics with some concrete alternative. As a matter of fact, the gravitational self-interacting potential is negligibly small compared to the usual external ones, making any resulting deviation, e.g., in the energy spectrum, extremely difficult to trial~\cite{carlip08, giulini11, meter11, yang13, grossardt16, gan16, bassi16, silva23}. Thus, instead of looking for stationary solutions of the Schr{\"o}dinger-Newton equation, one may use the particle spin as a witness of the gravitational self-interaction~\cite{hatifi20, grossardt21, grossardt23} in a Stern-Gerlach-like experiment where the only relevant potential will turn out to be the self-interacting one. By {\em ``Stern-Gerlach-like experiment"}, we mean a double Stern-Gerlach experiment followed by a spin measurement. This proposal is theoretically simple although experimentally challenging.

The paper is organized as follows. In Sec.~\ref{sec2}, we introduce the Schr{\"o}dinger-Newton equation for quantum particles. In Sec.~\ref{sec3}, we analyze the self-interaction effect in a Stern-Gerlach-like experiment. In Sec.~\ref{sec4}, we compare our results with those of Ref.~\cite{grossardt23} assuming physical parameters and make explicit that they complement each other. In Sec.~\ref{sec5}, we present our conclusions.

\section{The Schr{\"o}dinger-Newton equation for quantum particles}\label{sec2}

According to Di{\'o}si~\cite{diosi84}, the wave function~$\psi(\Vec{r}, t)$ of a nonrelativistic quantum particle with mass~$m$ would be evolved by the Schr{\"o}dinger-Newton equation
\begin{equation}
    i \hbar \frac{\partial}{\partial t} \psi(\Vec{r}, t)
    =
    \left(- \frac{\hbar^2}{2 m} \nabla^2 + V(\Vec{r}, t) + U(\Vec{r}, t)\right) \psi(\Vec{r}, t),
    \label{SNE}
\end{equation}
where~$V(\Vec{r}, t)$ is the usual external potential and
\begin{equation}
    U(\Vec{r}, t)
    \equiv
    - G m^2 \int d^3\Vec{r}\,' \, \frac{|\psi(\Vec{r}\,', t)|^2}{|\Vec{r} - \Vec{r}\,'|}
    \label{U}
\end{equation}
is the self-interacting potential. Here, $G$ is the gravitational constant. In our case, $\psi(\Vec{r}, t)$ will describe the center of mass of an extended object. In addition, we will consider a quantum state that will allow us to disregard the internal structure of the system in the self-interacting potential (in contrast to the case considered in Ref.~\cite{grossardt23}). The suitability of the {\em ``self-interacting"} qualifier for the potential~\eqref{U} comes from the fact that it depends on the particle state~$\psi(\Vec{r}, t)$ itself. While standard quantum mechanics asserts that~$|\psi(\Vec{r}, t)|^2$ is the probability density of finding the particle when (and only when) its position is measured, the presence of~$U(\Vec{r}, t)$ in Eq.~\eqref{SNE} implies that, to what concerns gravity, a quantum particle (yet point-like) would function as an extended system with an effective mass density given by
\begin{equation}
    \eta(\Vec{r}, t)
    \equiv
    m |\psi(\Vec{r}, t)|^2.
    \label{eta}
\end{equation}
Hence, yet tiny for elementary particles, the potential~\eqref{U} is at odds with standard quantum mechanics.

\section{The Schr{\"o}dinger-Newton equation in a Stern-Gerlach-like experiment}\label{sec3}

As mentioned below Eq.~\eqref{U}, we will consider a massive extended object (to maximize the effect of the gravitational self-interacting potential) with~$\psi(\Vec{r}, t)$ describing the center of mass of a spherically symmetric microcrystal with mass~$m$, radius~$R$, and spin~$s = 1/2$. Let us initially set the microcrystal in the normalized state 
\begin{equation}
    \Psi(\Vec{r}, 0)
    =
    \psi_\uparrow(\Vec{r}, 0) |\uparrow\,\rangle + \psi_\downarrow(\Vec{r}, 0) |\downarrow\,\rangle,
\end{equation}
centered at~$\Vec{r} = 0$, where
\begin{eqnarray}
    \psi_\uparrow(\Vec{r}, 0)
    &\equiv&
    G_\sigma(\Vec{r})^{1/2} \cos{\beta},
    \label{PSI1_0}
    \\
    \psi_\downarrow(\Vec{r}, 0)
    &\equiv&
    G_\sigma(\Vec{r})^{1/2} \sin{\beta},
    \label{PSI2_0}
\end{eqnarray}
$\beta \in (0, \, \pi/2)$, and
\begin{equation}
    G_\sigma(\Vec{r})
    =
    \left(\frac{1}{2 \pi \sigma^2}\right)^{3/2} \exp{\left(- \frac{|\Vec{r}|^2}{2 \sigma^2}\right)}
    \label{Gaussian}
\end{equation}
is the Gaussian distribution. Here, $\{|\uparrow\,\rangle, |\downarrow\,\rangle\}$ is the usual eigenstate basis of the $z$-axis spin operator~$\Hat{S}_z$:
\begin{equation}
    \Hat{S}_z |\uparrow\,\rangle
    \equiv
    (+ \hbar/2) |\uparrow\,\rangle,
    \quad
    \Hat{S}_z |\downarrow\,\rangle
    \equiv
    (- \hbar/2) |\downarrow\,\rangle.
\end{equation}
\begin{figure}
    \includegraphics[width=\columnwidth]{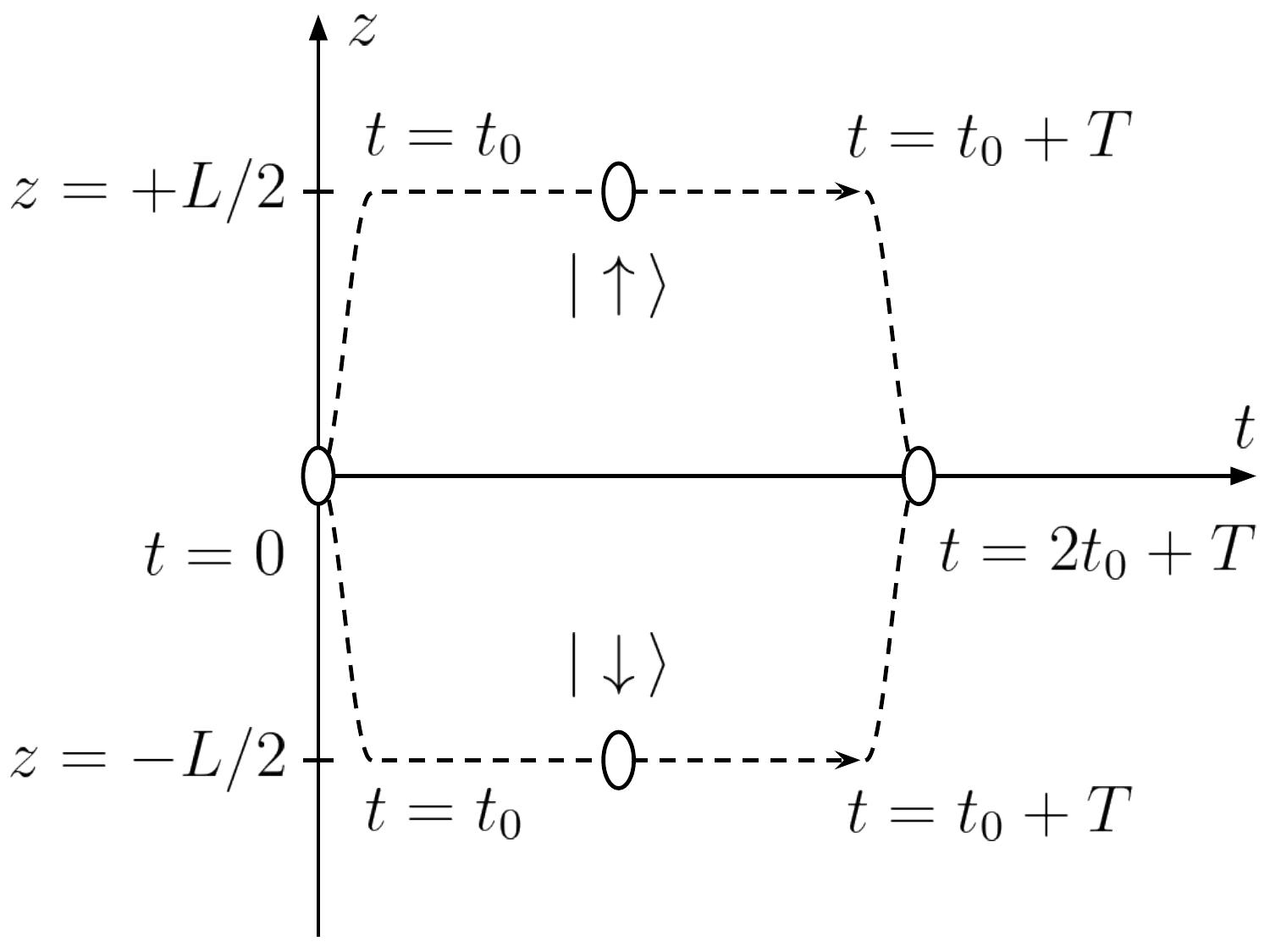}
    \caption{The system starts in the state~$\Psi(\Vec{r}, 0)$, transits to~$\Psi(\Vec{r}, t_0)$ when the Stern-Gerlach split is realized, evolves to~$\Psi(\Vec{r}, t_0 + T)$, and  ends up as~$\Psi(\Vec{r}, 2 t_0 + T)$ after the Stern-Gerlach split is reversed.}
    \label{fig1}
\end{figure}
By assuming that~$R \ll \sigma$, we are allowed to use Eqs.~\eqref{SNE}-\eqref{U} to evolve~$\Psi(\Vec{r}, t)$~\cite{diosi84, bassi16, giulini14}. We note that the case treated here is complementary to the one considered in Ref.~\cite{grossardt23}, where~$R \gg \sigma$ (as discussed in Sec.~\ref{sec4}).

Next, we realize a Stern-Gerlach split along the $z$~axis, driving~$\Psi(\Vec{r}, 0)$ into a spatial superposition with~$| \uparrow\,\rangle$ and~$|\downarrow\,\rangle$ lying apart by a distance~$L$ (see Fig.~\ref{fig1}):
\begin{equation}
    \Psi(\Vec{r}, t_0)
    =
    \psi_\uparrow(\Vec{r}, t_0) |\uparrow\,\rangle + \psi_\downarrow(\Vec{r}, t_0) |\downarrow\,\rangle,
\end{equation}
where (except for a common global phase)
\begin{eqnarray}
    \psi_\uparrow(\Vec{r}, t_0)
    &=&
    \psi_\uparrow(\Vec{r} - \Vec{r}_1, 0),
    \label{PSI1_t0}
    \\
    \psi_\downarrow(\Vec{r}, t_0)
    &=&
    \psi_\downarrow(\Vec{r} - \Vec{r}_2, 0),
    \label{PSI2_t0}
\end{eqnarray}
and 
\begin{equation}
    \Vec{r}_1
    \equiv
    (+ L/2) \, \Hat{z},
    \quad
    \Vec{r}_2
    \equiv
    (- L/2) \, \Hat{z}
\end{equation}
with $R \ll \sigma \ll L$. For this realization, one may employ an atom chip as a magnetic field source~\cite{margalit21}. In this way, it is possible to accelerate and decelerate the spin components~$\psi_\uparrow$ and~$\psi_\downarrow$ with a single device, alternating the direction of the electric current to designate the direction of the magnetic field applied to the microcrystal. Note that we have assumed~$t_0$ to be much smaller than the time interval~$T$ along which the system evolves free of external influences. Thus, we disregard self-interaction contributions in the time interval~$(0, \, t_0)$ compared to the interval~$(t_0, \, t_0 + T)$.

In the time interval~$(t_0, \, t_0 + T)$, the system evolves freely except for the self-interacting potential. In order to neglect the quantum spreading of the superposition components~$\psi_\uparrow(\Vec{r}, t)$ and~$\psi_\downarrow(\Vec{r}, t)$, let us begin writing the effective mass density associated with each one of them as [see Eq.~\eqref{eta}]
\begin{equation}
    \eta_{\uparrow\downarrow}(\Vec{r}, t)
    =
    m |\psi_{\uparrow\downarrow}(\Vec{r}, t)|^2
\end{equation}
and assign the corresponding masses
\begin{equation}
    m_{\uparrow\downarrow}(t)
    =
    \int d^3\Vec{r} \, \eta_{\uparrow\downarrow}(\Vec{r}, t).
\end{equation}
These components should spread at speeds of the order
\begin{equation}
    v_{\uparrow\downarrow}^{(\text{S})}
    \sim
    \frac{\hbar}{2 m_{\uparrow\downarrow} \sigma}.
    \label{vS}
\end{equation}
Equation~\eqref{vS} was estimated using Heisenberg's uncertainty principle: $\sigma \times (m_{\uparrow\downarrow} v_{\uparrow\downarrow}^{(\text{S})}) \sim \hbar/2$. Thus, in order to neglect the quantum spreading of~$\psi_\uparrow(\Vec{r}, t)$ and~$\psi_\downarrow(\Vec{r}, t)$, we must demand that~$v_{\uparrow\downarrow}^{(\text{S})} T \ll R$. In addition, we will consider physical situations where the gravitational effect of one superposition component on the other can be neglected, namely, that~$v_{\uparrow\downarrow}^{(\text{A})} T \ll R$, where
\begin{equation}
    v_{\uparrow\downarrow}^{(\text{A})}
    \sim
    a_{\uparrow\downarrow} T
    \sim
    \frac{G m_{\downarrow\uparrow} T}{L^2}.
\end{equation}

By dwelling in situations where~$T$ is small enough such that
\begin{equation}
    v_{\uparrow\downarrow}^{(\text{S})} T \ll R, 
    \quad
    v_{\uparrow\downarrow}^{(\text{A})} T \ll R,
    \label{displacement}
\end{equation}
the components $\psi_\uparrow(\Vec{r}, t)$ and~$\psi_\downarrow(\Vec{r}, t)$ will evolve according to Eq.~\eqref{SNE} with~$V(\vec{r}, t) = 0$ and the corresponding self-interacting potentials
\begin{eqnarray}
    U(\Vec{r}_1, t)
    &=&
    - G m^2 \int d^3\Vec{r}\,' \, \frac{|\psi_\uparrow(\Vec{r}\,', t)|^2}{|\Vec{r}_1 - \Vec{r}\,'|}
    \label{Ua1}
\end{eqnarray}
and
\begin{eqnarray}
    U(\Vec{r}_2, t)
    &=&
    - G m^2 \int d^3\Vec{r}\,' \, \frac{|\psi_\downarrow(\Vec{r}\,', t)|^2}{|\Vec{r}_2 - \Vec{r}\,'|},
    \label{Ua2}
\end{eqnarray}
respectively, where the potentials were evaluated at the center of the (peaked) distributions. In this case, $\psi_\uparrow(\Vec{r}, t)$ and~$\psi_\downarrow(\Vec{r}, t)$ barely change their spatial distribution:
\begin{equation}
    |\psi_{\uparrow\downarrow}(\Vec{r}, t_0)|^2
    \approx
    |\psi_{\uparrow\downarrow}(\Vec{r}, t_0 + T)|^2.
    \label{psiconst}
\end{equation}
Accordingly, the gravitational self-interacting potential turns out constant in the time interval~$(t_0, \, t_0 + T)$ [see Eqs.~\eqref{Ua1}-\eqref{Ua2}]:
\begin{equation}
    U(\Vec{r}_j, t_0)
    \approx
    U(\Vec{r}_j, t_0 + T)
    \quad
    (j = 1, 2).
\end{equation}

Moreover, we will constrain ourselves to situations where the kinetic energy
\begin{equation}
    K 
    \sim
    \sum_{\ell = \uparrow, \downarrow} \frac{1}{2} m_\ell \left({v_\ell^{(\text{S})}}^2 + {v_\ell^{(\text{A})}}^2\right)
\end{equation}
is much smaller than the absolute value of the self-interacting potential
\begin{equation}
    |U|
    \sim
    \sum_{\ell = \uparrow, \downarrow} \frac{G m_\ell^2}{\sigma},
\end{equation}
in which case~$\psi_{\uparrow\downarrow}(\Vec{r}, t)$ will be separately evolved by a simplified version of Eq.~\eqref{SNE}:
\begin{eqnarray}
    i \hbar \frac{\partial}{\partial t} \psi_{\uparrow}(\Vec{r}, t)
    &=&
    U(\Vec{r}_1, t_0) \psi_{\uparrow}(\Vec{r}, t),
    \label{SNEa1}
    \\
    i \hbar \frac{\partial}{\partial t} \psi_{\downarrow}(\Vec{r}, t)
    &=&
    U(\Vec{r}_2, t_0) \psi_{\downarrow}(\Vec{r}, t),
    \label{SNEa2}
\end{eqnarray}
and we recall that~$U(\Vec{r}_j, t)$ $(j = 1, 2)$ is constant in the interval~$(t_0, \, t_0 + T)$. It should be already clear at this point that the self-interacting potential will be responsible for causing a change of phase between~$\psi_\uparrow(\Vec{r}, t)$ and~$\psi_\downarrow(\Vec{r}, t)$, eventually.

The solutions of Eqs.~\eqref{SNEa1}-\eqref{SNEa2} are
\begin{eqnarray}
    \psi_\uparrow(\Vec{r}, t)
    &=&
    \psi_\uparrow(\Vec{r}, t_0) e^{i \varphi_\uparrow(\Vec{r}_1, t)},
    \label{PSIa1}
    \\
    \psi_\downarrow(\Vec{r}, t)
    &=&
    \psi_\downarrow(\Vec{r}, t_0) e^{i \varphi_\downarrow(\Vec{r}_2, t)},
    \label{PSIa2}
\end{eqnarray}
where
\begin{eqnarray}
    \frac{\partial}{\partial t} \varphi_\uparrow(\Vec{r}_1,t)
    &=&
    -  \frac{1}{\hbar}U(\Vec{r}_1, t_0),
    \label{SNEb1}
    \\
    \frac{\partial}{\partial t} \varphi_\downarrow(\Vec{r}_2, t)
    &=&
    - \frac{1}{\hbar}U(\Vec{r}_2, t_0),
    \label{SNEb2}
\end{eqnarray}
and the self-interacting potentials can be evaluated using Eqs.~\eqref{Ua1}-\eqref{Ua2}:
\begin{eqnarray}
    U(\Vec{r}_1, t_0)
    &=&
    - \sqrt{\frac{2}{\pi}} \frac{G m^2}{\sigma} \cos^2{\beta},
    \label{Ub1}
    \\
    U(\Vec{r}_2, t_0)
    &=&
    - \sqrt{\frac{2}{\pi}} \frac{G m^2}{\sigma} \sin^2{\beta}.
    \label{Ub2}
\end{eqnarray}
Then, by resolving  Eqs.~\eqref{SNEb1}-\eqref{SNEb2}, we can use Eqs.~\eqref{PSIa1}-\eqref{PSIa2} to write~$\psi_{\uparrow\downarrow}(\Vec{r}, t)$ at~$t = t_0 + T$ as (except for an arbitrary global phase)
\begin{eqnarray}
    \psi_\uparrow(\Vec{r}, t_0 + T)
    &=&
    \psi_\uparrow(\Vec{r}, t_0) e^{- i U(\Vec{r}_1, t_0) T/\hbar},
    \\
    \psi_\downarrow(\Vec{r}, t_0 + T)
    &=&
    \psi_\downarrow(\Vec{r}, t_0) e^{- i U(\Vec{r}_2, t_0) T/\hbar},
\end{eqnarray}
where~$\psi_\ell(\Vec{r}, t_0)$ ($\ell = \uparrow, \downarrow$) and~$U(\Vec{r}_j, t_0)$ ($j = 1, 2$) can be read from Eqs.~\eqref{PSI1_t0}-\eqref{PSI2_t0} and~\eqref{Ub1}-\eqref{Ub2}, respectively.

Finally, we reverse the Stern-Gerlach split at~$t = t_0 + T$ leading
\begin{equation}
    \Psi(\Vec{r}, t_0 + T)
    =
    \psi_\uparrow(\Vec{r}, t_0 + T) |\uparrow\,\rangle + \psi_\downarrow(\Vec{r}, t_0 + T) |\downarrow\,\rangle
\end{equation}
into
\begin{equation}
    \Psi(\Vec{r}, 2 t_0 + T)
    =
    \psi_\uparrow(\Vec{r}, 2 t_0 + T) |\uparrow\,\rangle 
    + 
    \psi_\downarrow(\Vec{r}, 2 t_0 + T) |\downarrow\,\rangle,
    \label{PSI_final}
\end{equation}
where (except for an irrelevant common global phase)
\begin{eqnarray}
    \psi_\uparrow(\Vec{r}, 2 t_0 + T)
    &=&
    \psi_\uparrow(\Vec{r}, 0) e^{- i U(\Vec{r}_1, t_0) T/\hbar},
    \\
    \psi_\downarrow(\Vec{r}, 2 t_0 + T)
    &=&
    \psi_\downarrow(\Vec{r}, 0) e^{- i U(\Vec{r}_2, t_0) T/\hbar},
\end{eqnarray}
and we recall that~$\psi_\ell(\Vec{r}, 0)$ ($\ell = \uparrow, \downarrow$) are given in Eqs.~\eqref{PSI1_0}-\eqref{PSI2_0}. For this reversion, one may employ in principle the same atom chip used in the Stern-Gerlach split. Nevertheless, it is worth noting that the (perfect) recombination of the superposition components is a challenging task~\cite{margalit21}. As previously discussed, this process occurs in a time interval~$t_0 \ll T$, allowing us to disregard self-interaction contributions in the interval~$(t_0 + T, \, 2 t_0 + T)$.

Now, let us express Eq.~\eqref{PSI_final} in the eigenstate basis~$\{|\rightarrow\,\rangle, |\leftarrow\,\rangle\}$ of the $x$-axis spin operator~$\Hat{S}_x$,
\begin{equation}
    \Hat{S}_x |\rightarrow\,\rangle
    \equiv
    (+ \hbar/2) |\rightarrow\,\rangle,
    \quad
    \Hat{S}_x |\leftarrow\,\rangle
    \equiv
    (- \hbar/2) |\leftarrow\,\rangle,
\end{equation}
as
\begin{equation}
    \Psi(\Vec{r}, 2 t_0 + T)
    =
    \psi_\rightarrow(\Vec{r}, 2 t_0 + T) |\rightarrow\,\rangle + \psi_\leftarrow(\Vec{r}, 2 t_0 + T)|\leftarrow\,\rangle
\end{equation}
with
\begin{equation}
    \psi_\rightarrow(\Vec{r}, 2 t_0 + T)
    =
    \frac{1}{\sqrt{2}} [\psi_\uparrow(\Vec{r}, 2 t_0 + T) + \psi_\downarrow(\Vec{r}, 2 t_0 + T)],
\end{equation}
\begin{equation}
    \psi_\leftarrow(\Vec{r}, 2 t_0 + T)
    =
    \frac{1}{\sqrt{2}} [\psi_\uparrow(\Vec{r}, 2 t_0 + T) - \psi_\downarrow(\Vec{r}, 2 t_0 + T)].
\end{equation}
Then, the probability of obtaining as an outcome~$+ \hbar/2$ in a measurement for the spin projection along the $x$~axis is
\begin{eqnarray}
    P_\rightarrow(\beta, T)
    &=&
    \int d^3\Vec{r} \, |\psi_\rightarrow(\Vec{r}, 2 t_0 + T)|^2
    \nonumber \\
    &=&
    \frac{1}{2} + \frac{1}{2} \sin{(2 \beta)} \cos{[\Delta\varphi(\beta, T)]},
    \label{P}
\end{eqnarray}
where, using Eqs.~\eqref{Ub1}-\eqref{Ub2}, we have
\begin{eqnarray}
    \Delta\varphi(\beta, T)
    &\equiv&
    - \frac{1}{\hbar} [U(\Vec{r}_1, t_0) - U(\Vec{r}_2, t_0)] T
    \nonumber \\
    &=&
    \sqrt{\frac{2}{\pi}} \frac{G m^2 \cos{(2 \beta)}}{\hbar \sigma} T.
    \label{DPa}
\end{eqnarray}
For this measurement, one may employ a second atom chip to realize a Stern-Gerlach split along the $x$~axis. Such an extra device does not compose the experiment discussed in Ref.~\cite{margalit21} and, therefore, may bring additional difficulties. Let us note that the usual quantum-mechanical result is recovered from the above making~$G = 0$:
\begin{equation}
    P_\rightarrow^\text{QM}(\beta)
    =
    \frac{1}{2} + \frac{1}{2} \sin{(2 \beta)}.
\end{equation}
In order to compare~$P_\rightarrow(\beta, T)$ with~$P_\rightarrow^\text{QM}(\beta)$, let us define
\begin{eqnarray}
    D_\rightarrow(\beta, T) 
    &\equiv& 
    P_\rightarrow(\beta, T) - P_\rightarrow^\text{QM}(\beta)
    \nonumber \\
    &=&
    - \sin{(2 \beta)} \sin^2{[\Delta\varphi(\beta, T)/2]}.
    \label{D}
\end{eqnarray}
To make explicit the challenge posed by the Planck scale in observing the change of phase induced by the self-interacting potential, let us express~$\Delta \varphi$ in Eq.~\eqref{DPa} as
\begin{equation}
    \Delta\varphi
    \approx
    2 \, \times \, 10^{14} \left(\frac{m}{m_\text{P}}\right)^2 \frac{(T/1 \, \si{\second})}{(\sigma/1 \, \si{\micro\meter})},
    \label{DPb}
\end{equation}
where $m_\text{P} \equiv \sqrt{\hbar c/G} \approx 2.2 \, \times \, 10^{-8} \, \si{\kilo\gram}$ is the reduced Planck mass and we have set~$\cos{(2 \beta)} \approx 1$. No self-interaction effect would be observable for~$\Delta\varphi \ll 1$. Fortunately, the smallness of~$m/m_\text{P}$ can be compensated by choosing an appropriate~$\sigma$ and a long enough spatial-superposition time~$T$, driving~$\Delta\varphi \sim 1$.

\section{Physical results}\label{sec4}

\begin{figure}
    \includegraphics[width=\columnwidth]{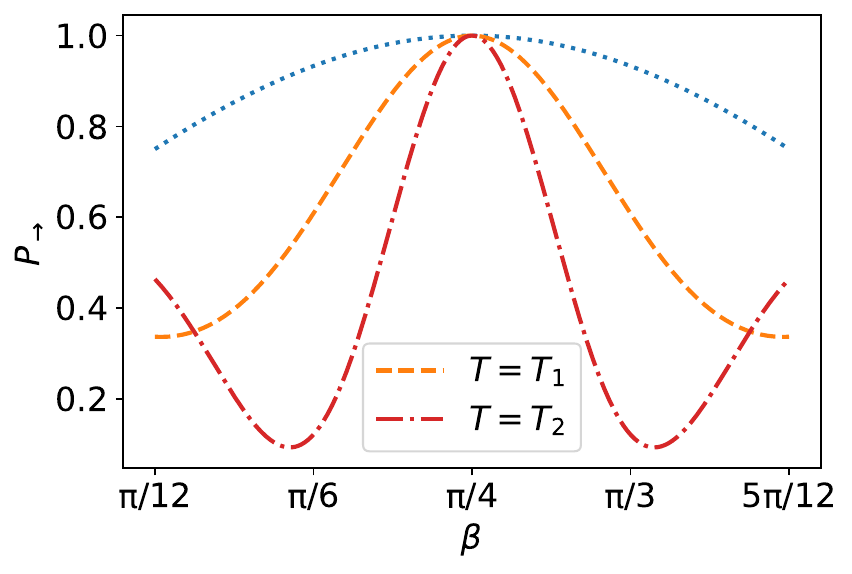}
    \caption{The dashed and dot-dashed lines are the plots of the probability~$P_\rightarrow(\beta, T)$ for~$T = T_1 = 2 \, \si{\second}$ and~$T = T_2 = 4 \, \si{\second}$, respectively. The dotted line represents the probability~$P_\rightarrow^\text{QM}(\beta)$ provided by standard quantum mechanics.}
    \label{fig2}
\end{figure}
\begin{figure}
    \includegraphics[width=\columnwidth]{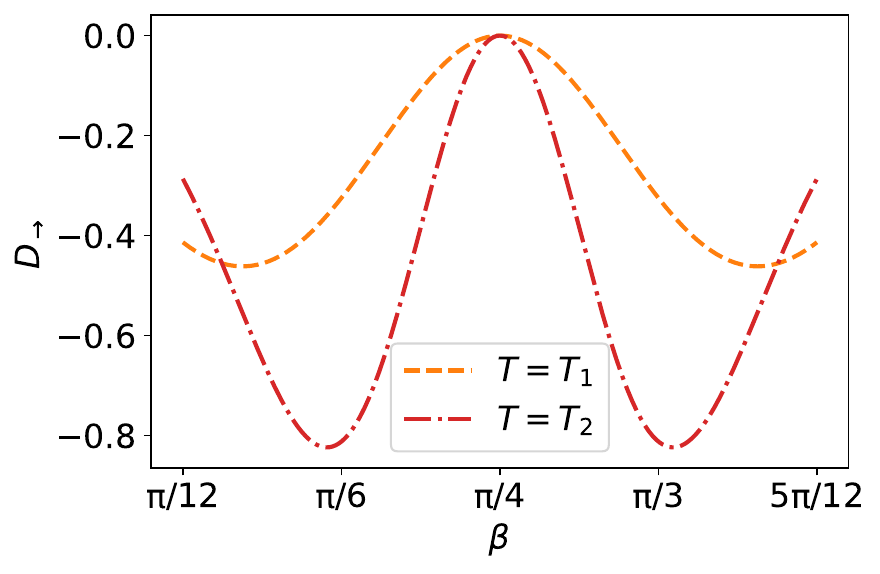}
    \caption{The dashed and dot-dashed lines are the plots of the probability difference~$D_\rightarrow(\beta, T)$ for~$T = T_1 = 2 \, \si{\second}$ and~$T = T_2 = 4 \, \si{\second}$, respectively.}
    \label{fig3}
\end{figure}
As suggested in Ref.~\cite{bose17}, let us consider a spherically symmetric homogeneous Ytterbium (Yb) microcrystal, with density~$\rho = 6.9 \, \times \, 10^{-15} \, \si{\kilogram/\micro\meter^3}$~\cite{haynes14}, doped with a single atom to give it a spin~$s = 1/2$. The mass of a microcrystal with radius~$R = 0.5 \, \si{\micro\meter}$ is~$m \approx 3.6 \, \times \, 10^{-15} \, \si{\kilo\gram}$.

The microcrystal will be initially described by the Gaussian distribution~\eqref{Gaussian} with standard deviation~$\sigma = 5 \, \si{\micro\meter}$, while the separation distance of the spatial superposition is chosen to be~$L = 100 \, \si{\micro\meter}$, satisfying the constraint~$L \gg \sigma \gg R$. The split time can be estimated to be of the order
\begin{equation}
    t_0 \sim \sqrt{\frac{2 m L}{g \mu_\text{B} \partial_z B}},
\end{equation}
where~$g$ is the electronic g-factor, $\mu_\text{B}$ is the Bohr magneton, and~$\partial_z B$ is the magnetic-field gradient in the $z$~direction~\cite{bose17}. For $\partial_z B \sim 1 \, \si{\tesla/\micro\meter}$, we obtain~$t_0 \sim 10^{-1} \, \si{\second}$, driving us to consider situations where~$T \gg 10^{-1} \si{\second}$. Notwithstanding, $T$ cannot be too large in order to protect the quantum state from interacting much with the environment. Then, we shall typically choose~$T \gtrsim 1 \si{\second}$.

Another point we must pay attention to is the satisfaction of Eq.~\eqref{displacement}. It is clear from Eq.~\eqref{vS} that the first constraint, $v_{\uparrow\downarrow}^{(\text{S})} T \ll R$, will not be satisfied if~$m_\uparrow$ or~$m_\downarrow$ is arbitrarily small. In order to avoid it, we must balance the values of~$m_\uparrow$ and~$m_\downarrow$. For this purpose, it is enough to  restrict the~$\beta \in (0, \, \pi/2)$ parameter [see below Eq.~\eqref{PSI2_0}] to~$\beta \in (\pi/12, \, 5 \pi/12)$. In this case,
\begin{equation}
    v_{\uparrow\downarrow}^{(\text{S})} T \sim 10^{-7} \, \si{\micro\meter}, 
    \quad
    v_{\uparrow\downarrow}^{(\text{A})} T \sim 10^{-10} \, \si{\micro\meter},
\end{equation}
respecting Eq.~\eqref{displacement} for~$R = 0.5 \, \si{\micro\meter}$. Likewise, the demand for the kinetic energy to be much smaller than the absolute value of the self-interacting potential is also satisfied:
\begin{equation}
  K/|U| \sim 10^{-9} \ll 1,
\end{equation}
vindicating Eqs.~\eqref{SNEa1}-\eqref{SNEa2}.

Figures~\ref{fig2} and~\ref{fig3} plot the probability~$P_\rightarrow(\beta, T)$ and the probability difference~$D_\rightarrow(\beta, T)$ as functions of~$\beta$, respectively. We see that~$D_\rightarrow(\pi/4, T) = 0$ since the superposition
\begin{eqnarray}
    \Psi(\Vec{r}, t_0)|_{\beta \to \pi/4}
    &\to&
    \frac{1}{\sqrt{2}} G_\sigma(\Vec{r} - \Vec{r}_1)^{1/2} |\uparrow\,\rangle
    \nonumber \\
    &+&
    \frac{1}{\sqrt{2}} G_\sigma(\Vec{r} - \Vec{r}_2)^{1/2} |\downarrow\,\rangle
\end{eqnarray}
is equally balanced, making the self-interacting potential have no net effect [see Eq~\eqref{DPa}]. The other local minima and maxima of~$D_\rightarrow(\beta, T)$ will depend on~$\Delta\varphi(\beta, T)$.
\begin{figure}
    \includegraphics[width=\columnwidth]{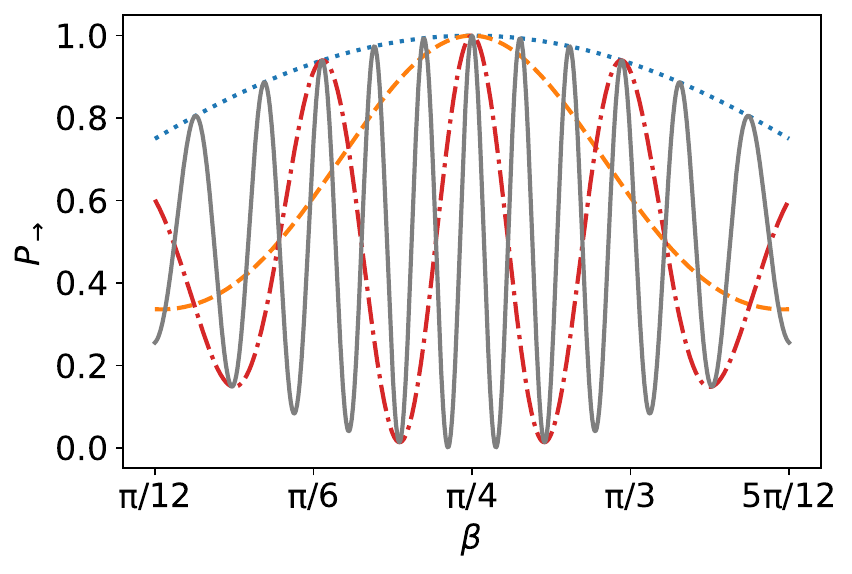}
    \caption{The probability~$P_\rightarrow(\beta, T)$ is plotted assuming a spherically symmetric Ytterbium microcrystal with radius~$R = 0.5 \, \si{\micro\meter}$ for~$T = 2 \, \si{\second}$. The dashed and dot-dashed lines obey Eq.~\eqref{P} with~$\sigma = 5 \, \si{\micro\meter}$ and~$\sigma = 1 \, \si{\micro\meter}$, respectively. The solid line is the corresponding plot according to Ref.~\cite{grossardt23}, where~$\sigma \ll R$. These curves should be compared against the dotted line representing the usual result provided by standard quantum mechanics. We see that our curves approach Gro{\ss}ardt's one as the value of~$\sigma$ gets smaller.}
    \label{fig4}
\end{figure}
\begin{figure}
    \includegraphics[width=\columnwidth]{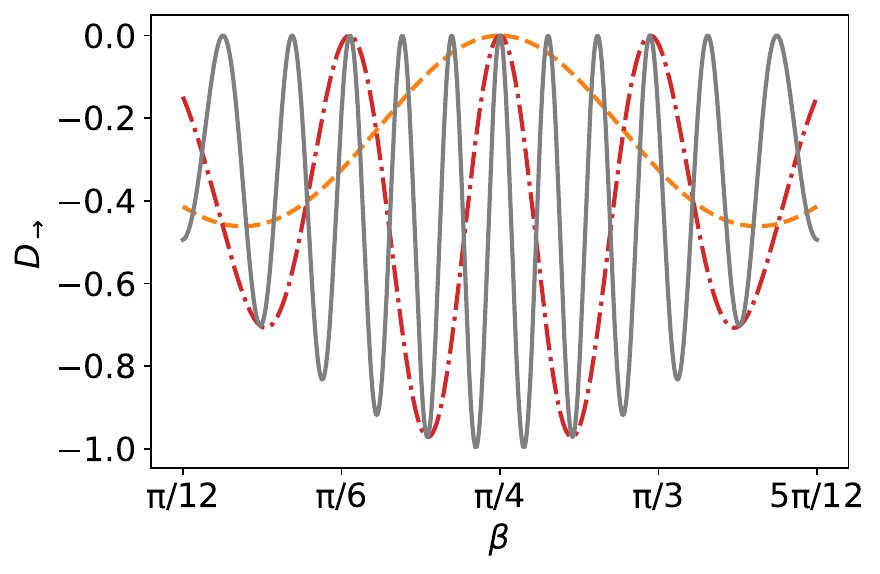}
    \caption{The probability difference~$D_\rightarrow(\beta, T)$ is plotted assuming a spherically symmetric Ytterbium microcrystal with radius~$R = 0.5 \, \si{\micro\meter}$ for~$T = 2 \, \si{\second}$. The dashed and dot-dashed lines obey Eq.~\eqref{D} with~$\sigma = 5 \, \si{\micro\meter}$ and~$\sigma = 1 \, \si{\micro\meter}$, respectively. The solid line is the corresponding plot according to Ref.~\cite{grossardt23}, where~$\sigma \ll R$. We see that our curves approach Gro{\ss}ardt's one as the value of~$\sigma$ gets smaller.} 
    \label{fig5}
\end{figure}

Our paper dwells in the regime~$\sigma \gg R$ which complements Ref.~\cite{grossardt23}, where~$\sigma \ll R$. In contrast to our case, where the change-of-phase formula~\eqref{DPa} depends on~$\sigma$, in Ref.~\cite{grossardt23} the internal structure of the microcrystal must be taken into account, and the change-of-phase formula has~$R$ in place of~$\sigma$ (multiplied by a constant of the order of the unity):
\begin{equation}
    \Delta\varphi(\beta, T)|_{\sigma \ll R}
    =
    \frac{6}{5} \frac{G m^2 \cos{(2 \beta)} }{\hbar R} T.
    \label{grossardt}
\end{equation}
It is clear, thus, that Eqs.~\eqref{DPa} and~\eqref{grossardt} should approach each other as~$\sigma$ approaches~$R$. This is made explicit in Figs.~\ref{fig4} and~\ref{fig5}. We see that the smaller the~$\sigma$ the more our curves for~$P_\rightarrow(\beta, T)$ and~$D_\rightarrow(\beta, T)$ plotted with Eq.~\eqref{DPa} ($\sigma > R)$ approach Gro{\ss}ardt's ones based on Eq.~\eqref{grossardt} ($\sigma < R)$. 

\section{Conclusions}\label{sec5}

Understanding why free macroscopic objects do not behave according to the predictions of quantum mechanics is an issue that remains elusive. A thought-provoking proposal is that a mechanism of gravitational self-interaction would clarify it~\cite{diosi84, penrose96, penrose98}. Nevertheless, such a proposal is intrinsically arduous to probe since usual external potentials overwhelm the gravitational self-interacting potential by many orders of magnitude for typical quantum particles. This drives deviations, e.g., in the energy spectrum due to the self-interacting potential tough to observe~\cite{carlip08, giulini11, meter11, yang13, grossardt16, gan16, bassi16, silva23}. To circumvent this difficulty, we have considered an experiment where the only potential is the self-interacting one. In this case, the challenge posed by the Planck scale can be compensated by choosing appropriate experimental parameters [see Eq.~\eqref{DPb}]. One should face the present paper and Ref.~\cite{grossardt23} as complementing each other, as explained above. Among the experimental challenges to probe our results are the necessity of suppressing channels of internal~\cite{japha23, henkel23} and environmental~\cite{isart11, schlosshauer19, kamp20} decoherence for a time interval of seconds. We hope this will be achieved in not too long~\cite{margalit21, bar-gill13, abobeih18, marshman22}.

\acknowledgments

The authors acknowledge discussions with Markus Aspelmeyer, Caslav Brukner, Nathan Argaman, Ron Folman, and Juan Pêgas. G.~H.~S.~A. was fully supported by the São Paulo Research Foundation (FAPESP) under grant~2022/08424-3. G.~E.~A.~M. was partially supported by the National Council for Scientific and Technological Development and FAPESP under grants~301508/2022-4 and~2022/10561-9, respectively.

\end{document}